\documentclass[aps,prl,twocolumn,groupedaddress,showpacs]{revtex4}

\usepackage{graphicx} 
\usepackage{amssymb} 

\usepackage{amsmath} 
\usepackage{dcolumn} 
\usepackage{bm}      

\begin{document}


\title{Nonclassical Rotational Inertia in Helium Crystals}


\author{A. C. Clark}
    \email{acc172@psu.edu}
\author{J. T. West}
\author{M. H. W. Chan}
\affiliation{Department of Physics, The Pennsylvania State
University, University Park, Pennsylvania 16802}


\date{\today}

\begin{abstract}

It has been proposed that the observed nonclassical rotational
inertia (NCRI) in solid helium results from the superflow of thin
liquid films along interconnected grain boundaries within the
sample. We have carried out new torsional oscillator measurements on
large helium crystals grown under constant temperature and pressure.
We observe NCRI in all samples, indicating that the phenomenon
cannot be explained by a superfluid film flowing along grain
boundaries.

\end{abstract}


\maketitle

The finding of NCRI in solid helium \cite{science} has been
replicated in torsional oscillator (TO) measurements in four other
laboratories \cite{RR,shkub,kojima}. The temperature dependence of
NCRI, characterized by saturation in the low temperature limit and a
gradual decay to zero at higher temperature, is qualitatively
reproducible in all measurements. However, the onset temperature
\textit{T$_O$}, the point where NCRI becomes resolvable from the
noise, varies between 150 mK and 400 mK in the studies of
commercially pure helium ($\sim$300 ppb of $^3$He). In addition,
relative to the total amount of $^4$He in the cell the NCRI fraction
(NCRIF) measured in the low temperature limit ranges from as little
as 0.03\% up to 20\% \cite{RR2}. In a large, cubic cell with linear
dimensions of $\sim$1 cm, Rittner and Reppy found the measured NCRIF
$\approx$ 0.5\% could be reduced to $<$ 0.05\% after annealing the
sample \cite{RR}. Although this appears consistent with two
numerical simulations in which perfect crystals are insulating
\cite{cepUmass}, less dramatic results have been observed in similar
annealing studies \cite{shkub,prl,ek3he}.

Although the presence of crystalline defects influences NCRIF,
neither the specific defects of importance, nor their relationship
with NCRI are known. Three fundamentally different kinds of defects
that are present in solid helium are point defects such as vacancies
or interstitials, dislocation lines, and grain boundaries.
Vacancies, which are likely more prevalent than interstitials, were
suggested to facilitate supersolidity in early theoretical
literature \cite{andchest}. Although recent investigations
\cite{umass1mahan} report attractive forces between them, a small
concentration of zero-point vacancies may still exist and facilitate
NCRI \cite{reatto}. However, it is impossible that they alone can
account for an NCRIF of 20\% \cite{RR2}.

There is an alternative model \cite{umass2balibar} in which NCRI
actually results from superfluid liquid $^4$He flowing along grain
boundaries. To be consistent with what is known of thin superfluid
films \cite{KT}, the $\sim$200 mK transition temperature implies an
effective thickness of 0.06 nm (one-fifth of a monolayer). Thus,
enormous surface areas of completely interconnected grain boundaries
are necessary to support a supercurrent constituting even just one
percent of the entire sample volume. This would require the
crystallites to have an average grain size of $\sim$20 nm. Even for
NCRIF = 0.03\% \cite{RR2,ek3he} the average grain size would be $<$
1 $\mu$m, whereas samples grown by the blocked capillary (BC) method
commonly result in crystals with linear dimensions $\geq$ 0.1 mm, as
determined by thermal conductivities \cite{therm} and x-ray
diffraction \cite{x}. In addition, annealing is found to increase
this value \cite{therm}. The same BC method was used in all previous
TO studies, and involves filling the sample cell through the hollow
torsion rod with high pressure liquid helium, freezing a section of
$^4$He in the filling line (the block), and then cooling the
constant volume below the block along the solid-liquid coexistence
boundary until solidification is complete.

Two growth techniques that are superior to the BC method
\cite{vos,heybey} are carried out at a fixed point anywhere on the
solid-liquid coexistence curve. The first, constant pressure (CP)
growth, is achieved by slowly cooling the cell while a specific
freezing pressure \textit{P$_F$} is maintained. The second, constant
temperature (CT) growth, takes place at a single freezing
temperature \textit{T$_F$} for minimal overpressures above
\textit{P$_F$}. The latter technique is most often (as it is in this
study) employed when growing low pressure solids from the superfluid
phase \cite{vos,heybey}. The first extensive investigations
demonstrating that large single crystals are reliably grown at CP/CT
combined either \textit{in situ} x-ray diffraction \cite{greywall}
or optical birefringence \cite{wanner,lee} techniques with sound
velocity measurements. An important experimental detail from Ref.'s
\cite{heybey,greywall,wanner,lee} is that the $^4$He filling line
necessarily remained open during solidification. Also, a cold
surface at one end of the cell seeded crystals, whereas the
remaining surfaces were poor thermal conductors so as to avoid the
nucleation of multiple crystallites. Due to such reliability, very
similar methods have been incorporated in all studies of $^4$He
single crystals since the early 1970s.

\begin{figure}[t]
\includegraphics[width=1.0\columnwidth]{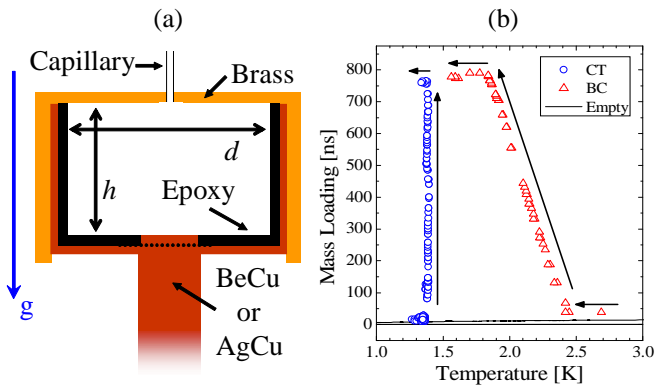}
\caption{\label{fig:one}(a) Schematic drawing of TO's. The epoxy
coating is 0.025 cm thick on the bottom of each cell, and 0.05/0.04
cm thick on the BeCu/AgCu walls. The 0.25 cm diameter BeCu cold spot
is recessed (to the dotted line) from the epoxy layer. In the AgCu
TO the cold surface is flush with the epoxy layer. Some relevant
parameters for the BeCu/AgCu TO are \textit{h} = 0.483/0.597 cm,
\textit{d} = 0.914/1.016 cm, $\tau$ = 0.933/1.27 ms (resonant period
at 20 mK). (b) Mass loading of BeCu TO during growth of two samples.
The minimum loading (solid at 25 bar) of the BeCu/AgCu TO is
740/2000 ns.}
\end{figure}

The motivation of the present work was to carry out a definitive
experiment to test the grain boundary model. In an effort to
separate the effects of isotopic impurities \cite{ek3he}, samples
were grown from both ``isotopically pure'' $^4$He ($\sim$1 ppb of
$^3$He) and commercially pure $^4$He ($\sim$300 ppb of $^3$He).
Measurements have been carried out in two torsional oscillators, one
made from beryllium copper (BeCu) and one from coin silver (AgCu)
(see Fig.~\ref{fig:one}a). Unlike other TO's in which a hollow
torsion rod serves as the $^4$He filling line
\cite{science,RR,shkub,kojima,RR2,prl,ek3he}, the torsion rod is
solid and a CuNi capillary (o.d. = 0.3 mm, i.d. = 0.1 mm) was
soldered to the opposite end of the cell. Apart from a small cold
spot from which to seed crystals, the walls are coated with a thin
layer of epoxy. This design allows us to seed the crystal at the
cell bottom and maintain an open filling line during freezing, and
thus enabled the growth of crystals at CT/CP within a TO for the
first time. Having replicated the precise conditions outlined in
Ref.'s \cite{heybey,greywall,wanner,lee}, we can be confident that
many of our samples are single crystals or at worst comprised of
just a few large crystals in the sample cell. We note that the
capillary can also be intentionally blocked with solid $^4$He during
the growth of a crystal, mimicking the BC method employed in
previous TO studies.

The mass loading of the BeCu TO during crystal formation is
displayed in Fig.~\ref{fig:one}b for two representative samples, one
grown at CT and one by BC. For the BC sample the molar volume
\textit{V$_M$} of the solid during growth ranges from 19.3
cm$^3$$\,$mol$^{-1}$ (at \textit{T} = 2.45 K and \textit{P} = 55
bar) to 20.7 cm$^3$$\,$mol$^{-1}$ (\textit{T} = 1.8 K and \textit{P}
= 30.7 bar). In contrast to this 7\% change in \textit{V$_M$}, the
sample grown at CT from the superfluid was subjected to variations
in \textit{V$_M$} of $\sim$0.07\% (i.e., temperature variations of
$\sim$20 mK).

\begin{figure}[t]
\includegraphics[width=1.0\columnwidth]{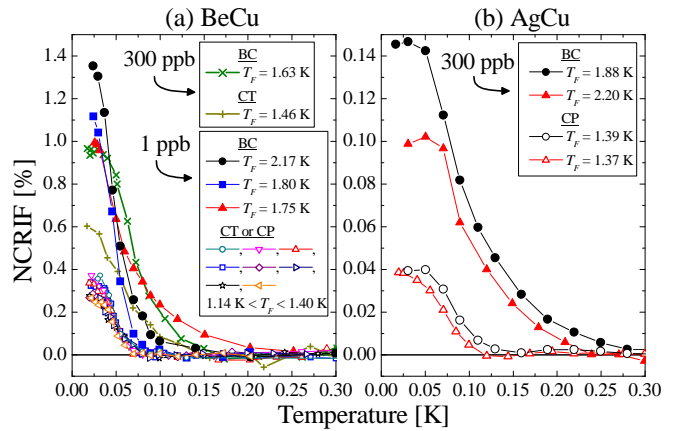}
\caption{\label{fig:two}(a) Comparison of several BC, CP, and CT
samples grown in the BeCu TO with two different impurity levels.
\textit{T$_F$} is the temperature where solidification is complete.
For eight 1 ppb crystals grown at CT/CP, 0.26\% $\leq$ NCRIF $\leq$
0.38\% ($\pm$0.01\%) and \textit{T$_O$} = 79 $\pm$ 5 mK. Among 12 BC
samples, 0.46\% $\leq$ NCRIF $\leq$ 2.0\% and 80 mK $\leq$
\textit{T$_O$} $\leq$ 275 mK. The two 300 ppb samples grown in the
BeCu TO are also shown. The maximum speed at the rim of the cell is
$<$ 5 $\mu$m$\,$s$^{-1}$ in all cases. (b) The four 300 ppb samples
grown in the AgCu TO. Rim speeds are $<$ 8 $\mu$m$\,$s$^{-1}$.}
\end{figure}

The temperature dependence of NCRIF in a number of BC and CT/CP
samples is shown in Fig.~\ref{fig:two}. For a particular cell the
NCRIF in BC samples is larger than that in samples grown at CT/CP.
This is true for both $^3$He concentrations. Further, we find that
\textit{T$_O$} is reduced when employing the CT/CP growth.
Surprisingly, there is an order of magnitude difference in the NCRIF
measured in the two cells. It appears that for both BC and CT/CP
samples, NCRIF is very sensitive to the exact internal geometry,
construction materials, and thermal properties of the cells. For
example, temperature gradients within the BeCu TO during growth are
larger than in the AgCu TO due to the thicker epoxy layer and the
much lower thermal conductivity of BeCu.

The most striking result from our study is that all eight 1 ppb
samples that were grown at CT/CP (at a rate of $\sim$1
$\mu$m$\,$s$^{-1}$) collapse onto a single curve above 40 mK and
thus share a common onset temperature. The vast improvement in
reproducibility over that of BC samples is most likely due to the
formation of single crystals within the cell. The spread in NCRIF at
lower temperatures may be related to differences in crystalline
defect (e.g., dislocations) densities.

\begin{figure}[t]
\includegraphics[width=1.0\columnwidth]{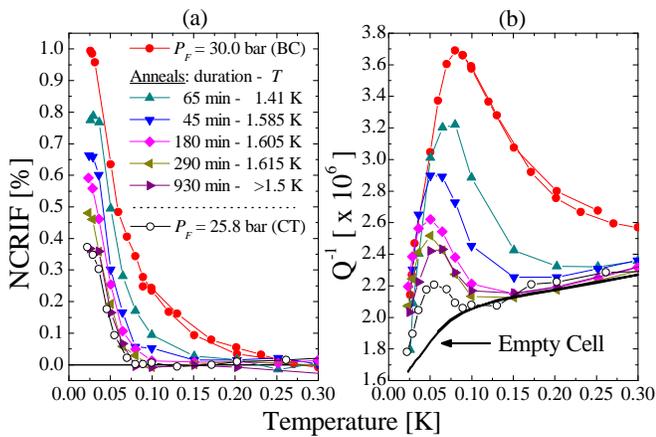}
\caption{\label{fig:three}(a) NCRIF after each sequential anneal for
a BC sample quenched from the bcc phase. A CT crystal grown from the
superfluid is potted for comparison. (b) Asymmetric reduction in
\textit{Q}$^{-1}$ following each anneal. All data were obtained with
rim speeds $<$ 3 $\mu$m$\,$s$^{-1}$.}
\end{figure}

We have carried out several annealing studies to investigate the
defects in our crystals. For a BC sample (see Fig.~\ref{fig:two}b,
\textit{T$_F$} = 2.20 K) in the AgCu TO we found NCRIF to
interestingly increase from 0.1\% to 0.2\% upon repeated annealing.
The dissipation, which accompanies NCRI in the form of a peak
\cite{science}, also increased. In the BeCu TO we found NCRIF in BC
samples to decrease with annealing. The most dramatic reduction
occurred in a 1 ppb sample grown with the BC method through the
bcc-hcp phase boundary (see Fig.~\ref{fig:three}). The sizeable tail
of NCRI is such that \textit{T$_O$} $\approx$ 275 mK. This was
dramatically reduced following the first anneal at relatively low
temperature. In fact, after 25 cumulative hours of annealing NCRIF
asymptotically approaches that found in CT/CP samples. For a 1 ppb
crystal freshly grown at CT there is no noticeable change in NCRIF
even after 40 h of annealing.

Annealing the BC sample in Fig.~\ref{fig:three} also led to a
dramatic reduction in the dissipation ($\propto$ \textit{Q}$^{-1}$).
Repeated heat treatments reduce the width of the peak, such that its
position remains close to the temperature where NCRIF changes most
rapidly. The fully annealed dissipation peak, just as NCRIF,
approaches that found in CT/CP samples. A phenomenological model
\cite{huse} associates the dissipation with a temperature dependent
coupling between the superfluid and normal components of the solid,
and predicts ($|\Delta\tau|$/$\tau$)/($\Delta$\textit{Q}$^{-1}$)
$\approx$ 1 for a homogeneous sample ($>$ 1 indicates
inhomogeneity). The ratio for this sample at different stages of
annealing evolves nonmonotonically from 9.5 to 12 to 10.5.

When the results of the present set of measurements are considered
together with the myriad of data from earlier studies
\cite{science,RR,shkub,kojima,RR2,prl,ek3he}, dislocations emerge as
a likely important class of defects. Dislocation lines form an
entangled web throughout each crystal, and can vary in density by
more than five orders of magnitude ($<$ 10$^5$ cm$^{-2}$ to
10$^{10}$ cm$^{-2}$) in solid helium samples grown above 1 K using
different methods. The actual line density deduced from sound
measurements \cite{dislocations} depends very sensitively (varying
by four orders of magnitude) on the exact growth conditions of
crystals \cite{dislocations,hiki}. It also can vary by at least one
order of magnitude from cell to cell, despite nearly identical
growth procedures \cite{dislocations,moredislocations}. The large
range of line densities and their sensitivity to sample growth and
containment can conceivably explain the very different NCRIF's
observed, even for single crystals. It is also known that only some
types of dislocations can be annealed away, which may explain the
unreliable effectiveness of annealing on the reduction of NCRIF.

The quantum mechanical motion of a single dislocation has recently
been considered \cite{degennes}, but a meaningful comparison with
experiments requires a thorough analysis of complex dislocation
networks. Recent simulations have shown that the core of some
dislocations may support superflow \cite{umass5}. Applying Luttinger
liquid theory to the dislocation network, Ref. \cite{umass5}
predicts that if the dislocation cores alone are responsible for
supersolidity then \textit{T$_C$} $\propto$
($\rho_S$/$\rho$)$^{0.5}$. Upon comparing several samples in
Fig.~\ref{fig:two}, the above relation is found not to be satisfied
if we take \textit{T$_C$} = \textit{T$_O$} and $\rho_S$/$\rho$ =
NCRIF. It has also been proposed that $^3$He impurities nucleate and
stabilize dislocations, which in turn produce a disordered
supersolid phase by providing a flow path for $^4$He interstitials
\cite{epl}.

\begin{figure}[t]
\includegraphics[width=1.0\columnwidth]{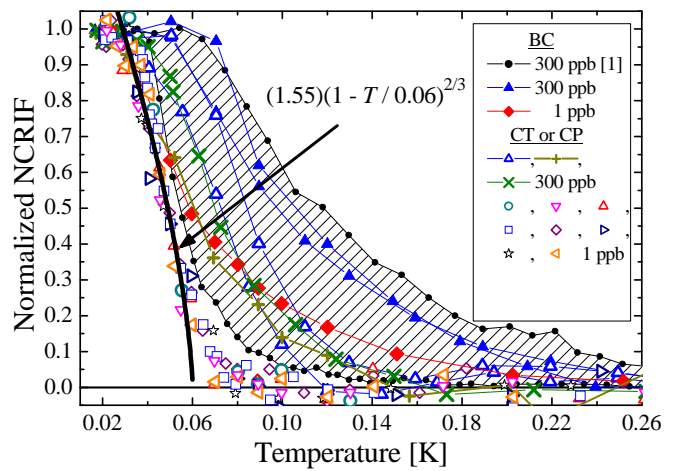}
\caption{\label{fig:four}Normalized NCRIF in various samples. There
is a wide spread in the data from the original KC experiment
\cite{science}. BC samples presented in Fig.~\ref{fig:two} possess a
high temperature tail of NCRI. CT/CP samples have a considerably
sharper onset. A two-thirds power law is plotted for comparison.}
\end{figure}

In addition to the magnitude of NCRIF, the high temperature tail and
thus \textit{T$_O$} are correlated with the way samples are prepared
(see Fig.~\ref{fig:two}). Several points can be drawn from the data
in Fig.~\ref{fig:four}, which are scaled by the low temperature
NCRIF. First, the high temperature tail of NCRI varies greatly for
different BC samples, which are presumably polycrystalline. Second,
the behavior of NCRIF in CT/CP samples is distinct in that the
temperature dependence is much sharper, with a well-defined onset
temperature. This is most apparent in crystals of 1 ppb purity. The
addition of $^3$He broadens the transition and pushes the onset of
NCRI to higher temperature. The sensitivity to impurities confirms
the general trends observed for a large number of BC samples that
were studied over a wide range of $^3$He concentrations
\cite{ek3he}.

All the data in Fig.~\ref{fig:four} were obtained at similar
measurement frequencies. A recent TO measurement \cite{kojima} on
the same 300 ppb sample at two different frequencies found
\textit{T$_O$} $\approx$ 250 mK at 1173 Hz and \textit{T$_O$}
$\approx$ 150 mK at 496 Hz. They also report irreversible changes in
NCRIF below $\sim$40 mK upon variation of the oscillation speed. We
have investigated \cite{TC} the thermal history of what appears to
be the same phenomenon in 1 ppb crystals and find that in the low
temperature limit there are in fact many metastable NCRIF's
available to the system. These results, as well as previously
observed critical velocities on the order of one quantum of
circulation \cite{prl}, indicate that the likely excitations in the
system are vortices.

A recent model \cite{anderson} capturing several aspects of the
experiments equates NCRI to the rotational susceptibility of a
vortex liquid phase. The high temperature tail of NCRI is said to
reflect the finite response time of vortices in the sample, which
are further slowed by $^3$He atoms dragged along with them. However,
the low temperature behavior of NCRIF \cite{kojima,TC} suggests that
at least a portion of the vortices are pinned. Trace amounts of
$^3$He are found to produce significant changes in the sound
velocity and elastic constants measured in solid $^4$He. These
results indicate that $^3$He impurities condense onto dislocation
lines \cite{iwasaPBD}. It is reasonable to assume that these
$^3$He-rich regions are the vortex pinning sites within the sample.
In this scenario the interplay between vortices, impurities, and
dislocations greatly impact the measured NCRIF in the solid.

A broad heat capacity peak near 75 mK was recently detected in solid
$^4$He \cite{xi}. This finding supports the notion that the
appearance of NCRI is a genuine signature of the transition between
the normal and supersolid phases. It is then natural to wonder if it
falls into the same universality class as that of superfluid $^4$He,
i.e., the 3D XY model. The gradual onset of NCRI previously observed
is not consistent with this expectation. However, the sharp onset
present in CT/CP crystals of 1 ppb purity is intriguing. We noted
above that the NCRIF's in all eight samples collapse onto a single
curve for \textit{T} $>$ 40 mK (see Fig.~\ref{fig:two}). The NCRIF
data of these crystals between 30 mK and 57 mK, as shown in
Fig.~\ref{fig:four}, can be represented by the expected two-thirds
power law, with a critical temperature \textit{T$_C$} $\approx$ 60
mK. If this highly speculative ``fit'' is applicable then the tail
between 60 mK and 79 mK is attributable to the finite measurement
frequency, residual (1 ppb) $^3$He impurities, and crystalline
defects. Measurements at a much lower frequency may reveal if there
is any validity to this speculation.

To summarize, we have shown that NCRI is found in large crystals of
solid $^4$He grown at constant CT/CP. In contrast to the results
from BC samples, the temperature dependence of NCRI in what are very
likely single crystals is reproducible and exhibits a sharp onset.

\begin{acknowledgments}
We thank P. W. Anderson, J. Beamish, W. F. Brinkman, D. A. Huse, E.
Kim, H. Kojima, X. Lin, N. Mulders, J. D. Reppy, and A. S. C.
Rittner for informative discussions. We also thank J. A. Lipa for
the 1 ppb purity helium and J. D. Reppy for the thin capillary used
in our cells. Support was provided by NSF Grants DMR 0207071 and
0706339.
\end{acknowledgments}


\begin{thebibliography}{31}
\bibitem{science} E. Kim and M. H. W. Chan, Science \textbf{305}, 1941 (2004).
\bibitem{RR} A. S. C. Rittner and J. D. Reppy, Phys. Rev. Lett. \textbf{97}, 165301 (2006).
\bibitem{shkub} A. Penzev, Y. Yasuta, and M. Kubota, J. Low Temp. Phys. \textbf{148}, 667 (2007); M. Kondo, S. Takada, Y. Shibayama, and K. Shirahama,
\textit{ibid.} \textbf{148}, 695 (2007).
\bibitem{kojima} Y. Aoki, J. C. Graves, and H. Kojima, Phys. Rev. Lett. \textbf{99}, 015301 (2007).
\bibitem{RR2} A. S. C. Rittner and J. D. Reppy, Phys. Rev. Lett. \textbf{98}, 175302 (2007).
\bibitem{cepUmass} D. M. Ceperley and B. Bernu, Phys. Rev. Lett. \textbf{93}, 155303
(2004); N. V. Prokof'ev and B. V. Svistunov, \textit{ibid.}
\textbf{94}, 155302 (2005).
\bibitem{prl} E. Kim and M. H. W. Chan, Phys. Rev. Lett. \textbf{97}, 115302 (2006).
\bibitem{ek3he} E. Kim, J. S. Xia, J. T. West, X. Lin, and M. H. W. Chan, Bull. Am. Phys. Soc. \textbf{52}, 610 (2007).
\bibitem{andchest} A. F. Andreev and I. M. Lifshitz, Sov. Phys. JETP \textbf{29}, 1107
(1969); G. V. Chester, Phys. Rev. A \textbf{2}, 256 (1970).
\bibitem{umass1mahan} M. Boninsegni \textit{et al.}, Phys. Rev. Lett. \textbf{97}, 080401 (2006); G. D. Mahan and H. Shin, Phys. Rev. B \textbf{74}, 214502 (2006).
\bibitem{reatto} D. E. Galli and L. Reatto, Phys. Rev. Lett. \textbf{96},
165301 (2006).
\bibitem{umass2balibar} E. Burovski, E. Kozik, A. Kuklov, N. V. Prokof'ev, and B. V. Svistunov, Phys. Rev. Lett. \textbf{94},
165301 (2005); S. Sasaki, R. Ishiguro, F. Caupin, H. J. Maris, and
S. Balibar, Science \textbf{313}, 1098 (2006); J. Low Temp. Phys.
\textbf{148}, 665 (2007); L. Pollet \textit{et al.}, Phys. Rev.
Lett. \textbf{98}, 135301 (2007).
\bibitem{KT} J. M. Kosterlitz and D. J. Thouless, J. Phys. C: Solid State Phys.
\textbf{6}, 1181 (1973); D. J. Bishop and J. D. Reppy, Phys. Rev.
Lett. \textbf{40}, 1727 (1978).
\bibitem{therm} F. J. Webb, K. R. Wilkinson, and J. Wilks, Proc. R. Soc. A \textbf{214}, 546 (1952); G. A. Armstrong, A. A. Helmy,
and A. S. Greenberg, Phys. Rev. B \textbf{20}, 1061 (1979).
\bibitem{x} A. F. Schuch and R. F. Mills, Phys. Rev. Lett. \textbf{8},
469 (1962).
\bibitem{vos} J. E. Vos \textit{et al.}, Physica \textbf{37}, 51 (1967).
\bibitem{heybey} O. W. Heybey and D. M. Lee, Phys. Rev. Lett. \textbf{19}, 106
(1967).
\bibitem{greywall} D. S. Greywall, Phys. Rev. A \textbf{3}, 2106 (1971).
\bibitem{wanner} R. Wanner and J. P. Franck, Phys. Rev. Lett.
\textbf{24}, 365 (1970).
\bibitem{lee} R. H. Crepeau, O. Heybey, D. M. Lee, and S. A. Strauss, Phys. Rev. A \textbf{3}, 1162 (1971).
\bibitem{huse} D. A. Huse and Z. U. Khandker, Phys. Rev. B \textbf{75}, 212504 (2007).
\bibitem{dislocations} R. Wanner, I. Iwasa, and S. Wales, Solid State Commun. \textbf{18}, 853 (1976).
\bibitem{hiki} F. Tsuruoka and Y. Hiki, Phys. Rev. B \textbf{20}, 2702 (1979).
\bibitem{moredislocations} I. Iwasa, K. Araki, and H. Suzuki, J. Phys. Soc. Jpn. \textbf{46}, 1119
(1979).
\bibitem{degennes} P.-G. de Gennes, C. R. Physique \textbf{7}, 561 (2006).
\bibitem{umass5} M. Boninsegni \textit{et al.}, arXiv:0705.2967v1 (2007).
\bibitem{epl} E. Manousakis, Europhys. Lett. \textbf{78}, 36002
(2007).
\bibitem{TC} A. C. Clark and M. H. W. Chan, to be published.
\bibitem{anderson} P. W. Anderson, Nature Physics \textbf{3}, 160 (2007).
\bibitem{iwasaPBD} H. Suzuki and I. Iwasa, J. Phys. Soc. Jpn. \textbf{49}, 1722
(1980); M. A. Paalanen, D. J. Bishop, and H. W. Dail, Phys. Rev.
Lett. \textbf{46}, 664 (1981).
\bibitem{xi} X. Lin, A. C. Clark, and M. H. W. Chan, to be published.
\end{thebibliography}

\end{document}